\documentclass{emulateapj}
\usepackage{apjfonts}

\newcommand{\Rsun}      {\mbox{$\rm\,R_{\mathord\odot}$}}

\submitted{Accepted by the Astrophysical Journal}

\begin{document}

\def\lsim{\mathrel{\lower .85ex\hbox{\rlap{$\sim$}\raise
.95ex\hbox{$<$} }}}
\def\gsim{\mathrel{\lower .80ex\hbox{\rlap{$\sim$}\raise
.90ex\hbox{$>$} }}}

\title{Identifications of Four {\em INTEGRAL} Sources in 
the Galactic Plane via {\em Chandra} Localizations}

\author{John A. Tomsick\altaffilmark{1},
Sylvain Chaty\altaffilmark{2},
Jerome Rodriguez\altaffilmark{2},
Luigi Foschini\altaffilmark{3},
Roland Walter\altaffilmark{4},
Philip Kaaret\altaffilmark{5}}

\altaffiltext{1}{Center for Astrophysics and Space Sciences, Code
0424, University of California at San Diego, La Jolla, CA,
92093, USA (e-mail: jtomsick@ucsd.edu)}

\altaffiltext{2}{AIM - Astrophysique Interactions Multi-\'echelles
(UMR 7158 CEA/CNRS/Universit\'e Paris 7 Denis Diderot),
CEA Saclay, DSM/DAPNIA/Service d'Astrophysique, B\^at. 709,
L'Orme des Merisiers, FR-91 191 Gif-sur-Yvette Cedex, France}

\altaffiltext{3}{INAF/IASF - Bologna, Via Gobetti 101, 
40129 Bologna, Italy}

\altaffiltext{4}{ISDC, Chemin d'Ecogia, 16, 1290 Versoix, Switzerland}

\altaffiltext{5}{Department of Physics and Astronomy, University of
Iowa, Iowa City, IA 52242, USA}

\begin{abstract}

Hard X-ray imaging of the Galactic plane by the {\em INTEGRAL}
satellite is uncovering large numbers of 20--100 keV ``IGR" 
sources.  We present results from {\em Chandra}, {\em INTEGRAL}, 
optical, and IR observations of 4 IGR sources:  3 sources in the 
Norma region of the Galaxy (IGR J16195--4945, IGR J16207--5129, 
and IGR J16167--4957) and one that is closer to the Galactic 
center (IGR J17195--4100).  In all 4 cases, one relatively 
bright {\em Chandra} source is seen in the {\em INTEGRAL} 
error circle, and these are likely to be the soft X-ray 
counterparts of the IGR sources.  They have hard 0.3--10 keV 
spectra with power-law photon indices of $\Gamma = 0.5$--1.1.  
While many previously studied IGR sources show high column 
densities ($N_{\rm H}\sim 10^{23-24}$ cm$^{-2}$), only 
IGR J16195--4945 has a column density that could be as high as 
$10^{23}$ cm$^{-2}$.  Using optical and IR sky survey catalogs 
and our own photometry, we have obtained identifications for all 
4 sources.  The $J$-band magnitudes are in the range 14.9--10.4, 
and we have used the optical/IR spectral energy distributions (SEDs) 
to constrain the nature of the sources.  Blackbody components 
with temperature lower limits of $>$9400 K for IGR J16195--4945 
and $>$18,000 K for IGR J16207--5129 indicate that these are
very likely High-Mass X-ray Binaries (HMXBs).  However, for 
IGR J16167--4957 and IGR J17195--4100, low extinction and the 
SEDs indicate later spectral types for the putative companions, 
indicating that these are not HMXBs.  

\end{abstract}


\section{Introduction}

The hard X-ray imaging of the Galactic Plane by the {\em INTErnational
Gamma-Ray Astrophysics Laboratory} \citep[{\em INTEGRAL},][]{winkler03}
is uncovering a large number of new or previously poorly studied high 
energy sources.  During the first two years of {\em INTEGRAL} operations 
(2002 October -- 2004 September), 56 new ``IGR" sources were discovered 
\citep{bird06}, and many more IGR sources have been discovered to date.  
{\em INTEGRAL} is also detecting many sources that are present in other 
X-ray catalogs but were not targets of focused studies until they were 
shown to be strong emitters of hard X-rays by {\em INTEGRAL}.  If 
these sources are included, well over 100 IGR sources have been 
found\footnote{A current list of IGR sources is available at
http://isdc.unige.ch/$\sim$rodrigue/html/igrsources.html}.  All or nearly
all of the IGR sources have been found in the 20--50 keV band with the 
Imager on Board the {\em INTEGRAL} Satellite (IBIS) coded aperture mask 
instrument \citep{ubertini03,lebrun03}.  Large numbers of IGR sources
have been found because of the combination of hard X-ray imaging with 
$12^{\prime}$ angular resolution, a large field of view ($9^{\circ}$ 
by $9^{\circ}$ fully coded FOV for IBIS), and the {\em INTEGRAL}
observing plan, which emphasizes observations of the Galactic Plane.

Follow-up observations of the IGR sources have shown a diversity
of source types, including Low-Mass X-ray Binaries (LMXBs), 
High-Mass X-ray Binaries (HMXBs), Active Galactic Nuclei, 
Cataclysmic Variables as well as other source types.  While
some of the sources have proved to be transient, others have
been consistently detected in X-rays in multiple {\em INTEGRAL}
observations as well as by other current and previous X-ray
instruments, suggesting that they are persistent.  One sub-group 
of IGR sources exhibits persistent but highly variable X-ray
emission.  These sources lie within a few degrees of the
Galactic Plane and show some evidence for clustering near Galactic 
spiral arms \citep{lutovinov05a}.  Their X-ray fluxes are typically 
$\sim$1--10 millicrab in the hard X-ray band \citep{dean05}, and 
their luminosities are mostly unknown due to large uncertainties
on their distances, but they may have luminosities of $10^{33-36}$ 
ergs~s$^{-1}$ if they are at typical 1--10 kpc Galactic distances.

Although X-ray, optical, and infrared (IR) observations of this 
group of persistent, Galactic IGR sources show that they do
not have uniform properties, some trends have been identified.
For many of these sources, their X-ray spectra show high column
densities with values of $N_{\rm H}$ well in excess of the levels
expected due to interstellar material.  The most extreme example
is IGR J16318--4848, for which $N_{\rm H}\sim 2\times 10^{24}$ 
cm$^{-2}$ \citep{mg03,walter03}, and there are
also several other sources with values of $N_{\rm H}$ in 
the $10^{23-24}$ cm$^{-2}$ range \citep[e.g., ][]{rodriguez03,
combi04,beckmann05}.  Many of the Galactic IGR sources also
exhibit X-ray pulsations, indicating the presence of a neutron
star.  Pulsations are detected for at least 5 of the persistent
IGR sources with periods of 139--1303~s 
\citep[e.g., ][]{lutovinov05b,bodaghee06}.  Finally, in some 
cases for which optical or IR spectra of IGR sources have been
obtained, high-mass stellar companions have been found.  
IGR J16318--4848 harbors an unusual supergiant B[e] star, and
the IR spectra also show P Cygni profiles, suggesting the 
presence of a strong stellar outflow \citep{fc04}.  Another
example is IGR J17391--3021 (= XTE J1739--302), which also
contains a supergiant \citep{smith06,negueruela06}.  

Here, we present results for X-ray, optical, and IR follow-up 
studies of 4 IGR sources that are in the Galactic Plane.  Three of 
the sources (IGR J16195--4945, IGR J16207--5129, IGR J16167--4957) 
are in the Norma region of the Galaxy (i.e., close to the tangent 
to the Norma spiral arm) at $332^{\circ} < l < 334^{\circ}$, and 
the fourth (IGR J17195--4100) is about half-way between the Norma 
region and the Galactic center at $l = 347^{\circ}$.  Although none 
of the sources have been previously well-studied, it has been pointed 
out that IGR J16195--4945 is probably the {\em Advanced Satellite 
for Cosmology and Astrophysics (ASCA)} source AX J161929--4945 
\citep{sugizaki01,sidoli05}.  The 2--10 keV {\em ASCA} spectrum of 
this source is well-described by a very hard power-law with a photon 
index of $\Gamma = 0.6^{+0.8}_{-0.5}$ and a relatively high column 
density of $N_{\rm H} = (12^{+8}_{-4})\times 10^{22}$ cm$^{-2}$.  
Based on its X-ray properties, it has been suggested that the source
is a neutron star HMXB \citep{sidoli05,bird06}.  IGR J16167--4957 
and IGR J17195--4100 have been tentatively identified with {\em ROSAT} 
sources \citep{stephen05}, but it appears that {\em INTEGRAL} is 
the first X-ray satellite to detect IGR J16207--5129.  The nature
of these 3 sources is completely uncertain \citep{bird06}.

In this work, we use observations with the {\em Chandra X-ray 
Observatory} \citep{weisskopf02} to obtain sub-arcsecond X-ray 
positions for the 4 sources (\S$2$) and to determine whether
the sources are associated with known sources (\S$3$).  In 
\S$4$ and \S$5$, we present a study of the {\em Chandra} and
{\em INTEGRAL} spectra.  In \S$6$, we report on optical and
IR photometry of these sources we made at ESO's New Technology
Telescope (NTT), and we combine these measurements with those
available in optical and IR catalogs in order to constrain the
optical and IR spectral energy distributions (\S$7$).  In 
\S$8$ and \S$9$, we discuss our results and present our 
conclusions.  

\section{{\em Chandra} Observations and Source Detection}

We obtained short, $\sim$5~ks, {\em Chandra} observations of
the fields of IGR J16195--4945, IGR J16207--5129, IGR J16167--4957, 
and IGR J17195--4100 with the primary goal of localizing the sources 
to facilitate IR and optical identifications.  We chose targets 
located close to the Galactic Plane that were detected in the 
20--40~keV band during our 2003 February {\em INTEGRAL} observation 
of the black hole transient 4U 1630--47
\citep{tomsick04_munich,tomsick05} and that were also reported in the 
first and second catalogs of {\em INTEGRAL} sources 
\citep{bird04,bird06}.  Consistent hard X-ray detections gave us 
reason to believe that these were persistent sources and that the 
probability of detection with {\em Chandra} was high.

Table~\ref{tab:obs} shows the Observation IDs (ObsIDs) and 
times of our {\em Chandra} observations of the 4 IGR sources, 
which took place between 2005 April and July.  We used the 
Advanced CCD Imaging Spectrometer \citep[ACIS,][]{garmire03} 
with the center of the $3^{\prime}$ {\em INTEGRAL} error 
circle placed at the nominal aimpoint for the ACIS-I array.  
Although the effective exposure times are lower in the gaps 
between ACIS-I chips, we used a dithering pattern with an 
amplitude twice as large as the standard pattern to achieve 
a more uniform sensitivity over the {\em INTEGRAL} error 
circle.  In processing the data, we started with the 
``level 1'' event list produced by the standard data 
processing with ASCDS versions between 7.5.3 and 7.6.2.  
For further processing, we used the {\em Chandra} Interactive 
Analysis of Observations (CIAO) version 3.3.0.1 software and 
Calibration Data Base (CALDB) version 3.2.1.  We used the 
CIAO routine {\tt acis\_process\_events} to obtain a 
``level 2'' event list and image.

\begin{figure}
\centerline{\includegraphics[width=0.50\textwidth]{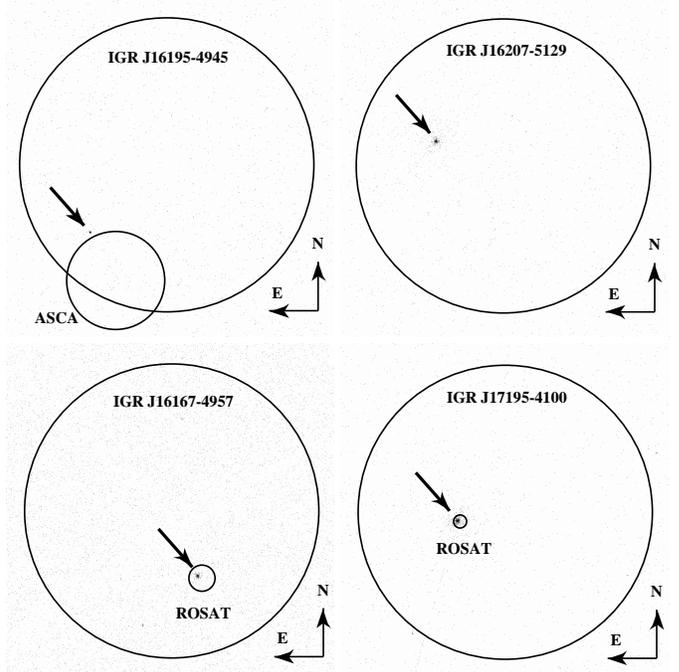}}
\caption{{\em Chandra} 0.3--10 keV images of the 4 IGR sources.
The large circles are the $3^{\prime}$ {\em INTEGRAL} error circles.
In three cases, {\em ASCA} or {\em ROSAT} counterparts have been
suggested, and their error circles are also shown.  Arrows point 
to the brightest {\em Chandra} source detected in each observation, 
and these are very likely the soft X-ray counterparts to the IGR 
sources.  The lengths of the N and E arrows are $1^{\prime}$.
\label{fig:images}}
\end{figure}

The {\em Chandra} images are shown in Figure~\ref{fig:images}.
For each target, we searched for sources in a $7\times 7$ 
arcmin$^{2}$ region that includes the full {\em INTEGRAL} 
error circle.  We restricted the energy range to 
0.3--10 keV and used the CIAO routine {\tt wavdetect} 
\citep{freeman02}.  Based on the image size of 854-by-854 
pixels and our detection threshold of $10^{-6}$
\citep[see][for a precise definition]{freeman02}, we would 
expect to detect $\lsim$1 spurious source.  For ObsIDs 5471, 
5472, 5473, and 5474, we detected 4, 7, 4, and 7 sources, 
respectively.  The brightest sources for these four ObsIDs 
have 183, 678, 866, and 676 counts, respectively, while the 
remaining sources have between 3 and 24 counts.  In addition, 
each of these four brightest sources are located within the 
respective {\em INTEGRAL} error circles (see Figure~\ref{fig:images}), 
making it very likely that these sources are, in fact, the soft 
X-ray counterparts of the IGR sources.  In the following, we focus 
on the properties of these four sources and primarily
refer to them by their IGR names (even though {\em Chandra} names
are also given to the sources).  

\section{{\em Chandra} Positions and Associations with Known Sources}

Table~\ref{tab:loc} provides the {\em Chandra} positions for 
the four sources.  The positions come from the {\tt wavdetect} 
analysis described above, and the position uncertainties are 
$0^{\prime\prime}.6$.  This value is the nominal {\em Chandra}
pointing accuracy, and the statistical errors calculated by
{\tt wavdetect} are negligible in comparison.  Here, we use the 
SIMBAD database to determine if the {\em Chandra} sources are 
associated with previously known sources.

For IGR J16195--4945, the {\em Chandra} source CXOU J161932.2--494430
lies $1^{\prime}.1$ from the best {\em ASCA} position for AX J161929--4945, 
which is just outside the $1^{\prime}$ {\em ASCA} error circle (see 
Figure~\ref{fig:images}).  While \cite{sugizaki01} quote {\em ASCA}
position uncertainties of $1^{\prime}$, a careful analysis by 
\cite{ueda99} shows that the 90\% confidence error radii for {\em ASCA}
sources are in the range $0^{\prime}.6$--$0^{\prime}.8$, depending on
the detection significance.  Thus, the $1^{\prime}$ error circle can 
be considered to be slightly larger than a 90\% confidence error 
circle.  Still, CXOU J161932.2--494430 and AX J161929--4945 are close 
enough to consider the association likely.  An association 
between IGR J16195--4945, AX J161929--4945, and the B1/B2~Ia 
supergiant HD~146628 has also been suggested 
\citep{sidoli05,tomsick04_munich,sugizaki01}.  However, the {\em Chandra} 
position is $1^{\prime}.2$ away from HD~146628, clearly ruling out 
this association.  There are no sources in the SIMBAD database within 
$1^{\prime}$ of the {\em Chandra} position.

There is no evidence that IGR J16207--5129 was previously 
detected in the X-ray band.  While the source is bright enough 
to be seen in the {\em ASCA} Galactic Plane Scans, it lies just 
outside of the region that was observed at a Galactic latitude 
of $b = -1.05^{\circ}$.  A possible association between 
IGR J16207--5129 and the A1~IVe star HD~146803 was suggested
\citep{tomsick04_munich,masetti06}; however, this is ruled out 
by the fact that the positions for HD~146803 and 
CXOU J162046.2--513006 are not compatible.  There are no sources 
in the SIMBAD database within $1^{\prime}$ of the {\em Chandra} 
position.

For IGR J16167--4957 and IGR J17195--4100, the {\em Chandra} 
positions confirm associations with 1RXS~J161637.2--495847 and 
1RXS~J171935.6--410054, respectively (see Figure~\ref{fig:images}).  
A SIMBAD search at the {\em Chandra} position of 
CXOU J161637.7--495844 reveals only IGR J16167--4957 
and the {\em ROSAT} source within $1^{\prime}$.  For IGR J17195--4100, 
the only other SIMBAD source within $1^{\prime}$ of CXOU J171935.8--410053 
is USNO-B1.0 0489--0511283.  This USNO source is $0^{\prime\prime}.70$ 
from the {\em Chandra} position and must be considered as a possible 
optical counterpart.  We investigate this possibility further in \S$6$.

\begin{figure}
\centerline{\includegraphics[width=0.50\textwidth]{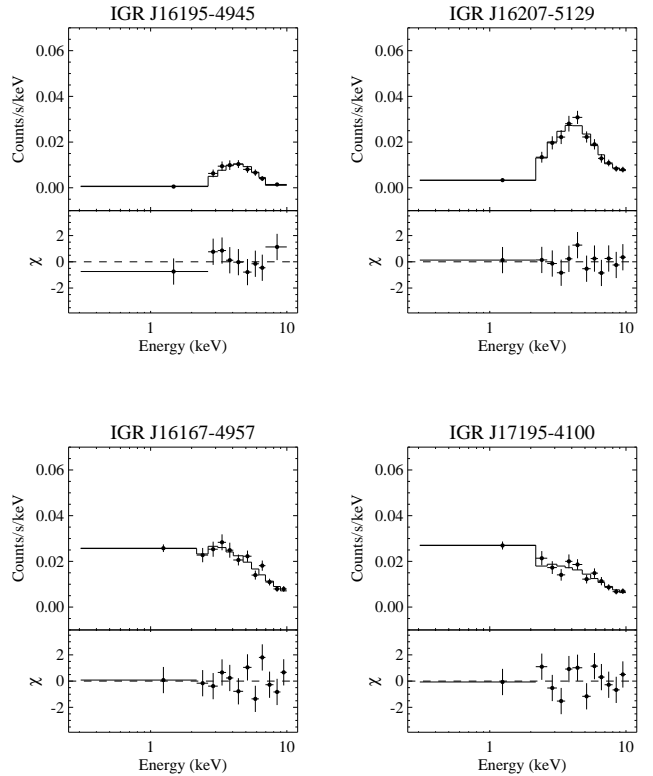}}
\caption{{\em Chandra} energy spectra and residuals for the four 
IGR sources fitted with an absorbed power-law model.  For 
IGR J16207--5129, IGR J16167--4957, and IGR J17195--4100, a
pile-up correction is included in the model.\label{fig:spectra}}
\end{figure}

\section{{\em Chandra} Energy Spectra and Light Curves}

We produced 0.3--10 keV energy spectra for each of the four bright 
{\em Chandra} sources using the CIAO tool {\tt dmextract} to produce 
the spectrum and the tools {\tt mkacisrmf} and {\tt mkarf} to 
produce the files needed to characterizing the instrument response.  
We used a circular source extraction region with a radius of 
$2^{\prime\prime}.5$ and an annular background region centered on 
the source with inner and outer radii of $10^{\prime\prime}$ 
and $60^{\prime\prime}$, respectively.  Given that two of the sources 
are close to the gaps between the ACIS chips, we produced exposure 
maps to determine the effective exposure time for each source.  The
sources IGR J16207--5129 and IGR J17195--4100 (ObsIDs 5472 and 5474) 
received the full exposure time, but IGR J16195--4945 (ObsID 5471) and 
IGR J16167-4957 (ObsID 5473) received 70\% and 67\% of the full 
exposure time, respectively.  The CIAO tool {\tt mkarf} automatically 
accounts for the reduced instrument response near chip boundaries.

We fitted the spectra using the software package XSPEC 11.3.2o. 
Originally, we rebinned the spectra and fitted them using
$\chi^{2}$-minimization.  After rebinning, the spectrum for the
lowest count rate source (IGR J16195--4945) had 9 bins with an 
average of 20 counts per bin, while we used 12-bin spectra for the 
other 3 sources.  We fitted the spectra using a model consisting of 
an absorbed power-law, and, based on the results for other IGR sources, 
it would not be surprising if the column density has intrinsic and 
interstellar contributions \citep[e.g.,][]{walter03,revnivtsev03a}.  
The absorbed power-law fit is acceptable ($\chi^{2}/\nu = 4.0/6$) 
for IGR J16195--4945, but we obtained very poor fits for the other 
three sources with values of $\chi^{2}$ between 46 and 83 for 9 degrees 
of freedom (dof).  Large positive residuals above $\sim$6~keV are the 
main reason for the poor fits.  This suggests that photon pile-up 
\citep{davis01} is affecting these three spectra, and, indeed, one 
expects pile-up to significantly impact the spectral shape for sources 
with count rates as high as we observe for the three brighter sources 
(0.13--0.18 s$^{-1}$).  We refitted the three affected spectra after
including the XSPEC {\tt pileup} model \citep{davis01}, and the
quality of the fits improves enormously with $\chi^{2}$ values
of 3.7, 8.7, and 9.2 (8 dof) for IGR J16207--5129, IGR J16167--4957, 
and IGR J17195--4100, respectively.  Comptonization models (e.g., 
{\tt comptt} in XSPEC) approximate a power-law shape in the 
0.3--10 keV band and provide fits of similar quality to the
power-law fits.  While a thermal blackbody model can adequately
describe the spectrum of the fainter source, such a model provides
very poor fits to the spectra of the three brighter sources.  Even 
if the pile-up correction is included, an absorbed blackbody model
gives $\chi^{2}$ values between 32 and 98 for 8 dof.  We conclude 
that these three spectra are non-thermal, and we focus on the power-law 
model below.

\begin{figure}
\centerline{\includegraphics[width=0.50\textwidth]{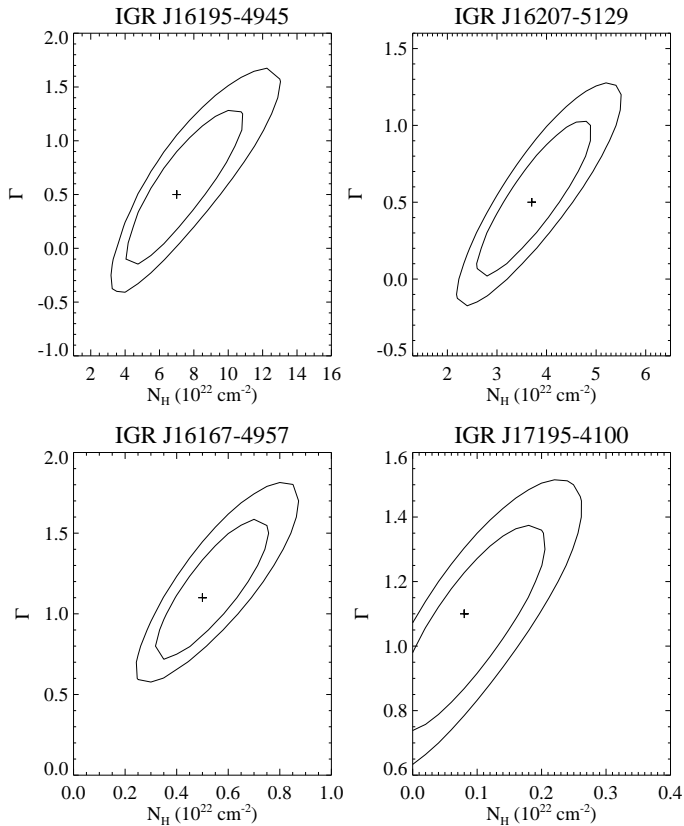}}
\caption{Error contours for the parameters $N_{\rm H}$ and $\Gamma$
from the fits to the {\em Chandra} spectra.  The inner contour is
68\% confidence ($\Delta$C = 2.3) and the outer contour is
90\% confidence ($\Delta$C = 4.6).\label{fig:contour}}
\end{figure}

Another signature of photon pile-up is the flattening of the point 
spread function (PSF), and this causes the three brighter sources
to appear extended when their radial profiles (i.e., surface
brightness vs.~angle from the source position) are compared to
the profiles of point sources that are unaffected by pile-up.
We studied the radial profiles of the 4 IGR sources as a check
of the pile-up correction, and we conclude that the PSF distortions 
are well-explained by pile-up only and that the pile-up corrections 
we apply in performing the spectral fits are correct.

Figure~\ref{fig:spectra} shows the 0.3--10 keV spectra and residuals 
for the absorbed power-law fits (with the pile-up correction), and 
Table~\ref{tab:spectra} provides the values of the spectral parameters.  
To obtain these values, we refitted the spectra by minimizing the Cash 
``C"-statistic \citep{cash79}, which is an improvement over 
$\chi^{2}$-minimization for spectra with small numbers of counts as 
the Cash analysis does not require discarding information by rebinning 
the spectra.  While none of the sources show the extremely high column 
densities that have been seen for some of the IGR sources, the measured 
values of $N_{\rm H}$ for IGR J16195--4945 and IGR J16207--5129 are 
$7^{+5}_{-3} \times 10^{22}$ and $3.7^{+1.4}_{-1.2} \times 10^{22}$ 
cm$^{-2}$ (90\% confidence errors), placing them somewhat above 
their respective Galactic column densities of $2.2\times 10^{22}$ and 
$1.7\times 10^{22}$ cm$^{-2}$ \citep{dl90}.  IGR J16167--4957 and 
IGR J17195--4100 have values of $N_{\rm H}$ that are significantly
lower than their Galactic column densities.  

All the sources are intrinsically hard in the 0.3--10 keV band based 
on the measured values of the power-law photon index.  As shown in 
Table~\ref{tab:spectra}, the best fit values of $\Gamma$ range from 
0.5 to 1.1, and, in all cases, the spectra are constrained so that 
$\Gamma < 1.6$ based on the 90\% confidence ($\Delta$$C = 2.7$) 
error bars.  While the errors on the individual parameters indicate 
very hard spectra, we also calculated error contours to account for 
possible correlations between $N_{\rm H}$ and $\Gamma$.  The contour 
plots in Figure~\ref{fig:contour} show the 68\% ($\Delta$$C = 2.3$) 
and 90\% ($\Delta$$C = 4.6$) confidence error contours, and it is 
clear that these parameters are significantly correlated for all 4
spectra.  While this analysis shows that the error regions are
somewhat larger than indicated by the individual parameter values
given in Table~\ref{tab:spectra}, the spectra are still constrained
to be quite hard, with $\Gamma < 1.8$ (90\% confidence), and it is 
still the case that IGR J16195--4945 and IGR J16207--5129 have
column densities that are higher than the Galactic values while
$N_{\rm H}$ is significantly less than the Galactic values for the
other two sources.

\begin{figure}
\centerline{\includegraphics[width=0.50\textwidth]{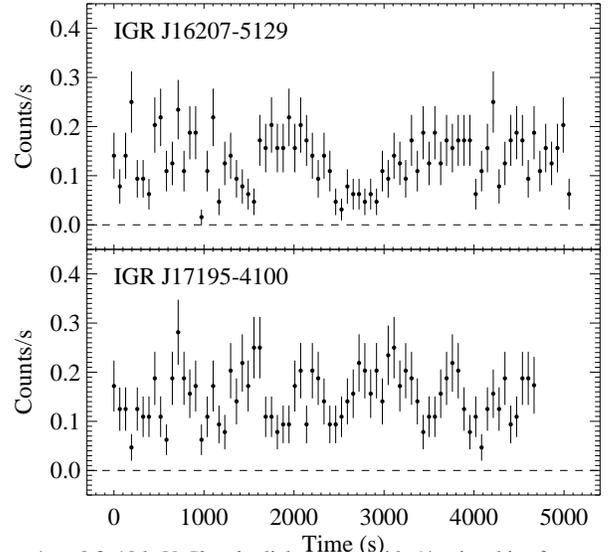}}
\caption{0.3--10 keV {\em Chandra} light curves with 64~s time 
bins for two of the IGR sources.  IGR J16207--5129 is clearly 
variable, and there is evidence that IGR J17195--4100 is also 
variable (see text).  The light curves for the other two IGR
sources are significantly affected by dithering over the gap
between ACIS chips and are not shown.\label{fig:lightcurves}}
\end{figure}

We produced 0.3--10~keV light curves with 64~s time bins for all 
4 sources.  The light curves for IGR J16195--4945 and IGR J16167--4957 
are strongly affected by the dithering over gaps between the ACIS chips.  
In both cases, the count rates drop to zero several times during the 
observation.  In all, about 30\% of the time bins are affected, and we
do not study the light curves for these 2 sources further.  On the 
other hand, there are no problems with the light curves for 
IGR J16207--5129 and IGR J17195--4100, and these are shown in 
Figure~\ref{fig:lightcurves}.  The clearest variability is seen for 
IGR J16207--5129.  In the middle of the observation, the count rate 
drops to about a third of its average level, remains there for 
$\sim$500~s, and then recovers.  We formally tested for variability
by fitting each light curve with a constant and obtained 
$\chi^{2}/\nu = 163/78$ and $\chi^{2}/\nu = 100/72$ for 
IGR J16207--5129 and IGR J17195--4100, respectively.  Thus, it is 
likely that both sources exhibit some variability.

\section{{\em INTEGRAL} Energy Spectra}

We also produced $>$22 keV spectra from the IBIS/ISGRI instrument on 
{\em INTEGRAL}. The observations used to produce the spectra include 
all public data available from {\em INTEGRAL} revolutions 46 to 235 (2003 
February 27 to 2004 September 15).  We made sky images and mosaics with 
the Off-line-Analysis Software version 5.1 in 10 energy bands between 
22 keV and 250 keV, and we derived average source spectra from the 
mosaic images.  After making vignetting, off-axis, and dead-time
corrections, the effective exposure time is 495~ks for IGR J17195--4100 
and $\sim$165~ks per source for the other 3 sources (see 
Table~\ref{tab:integral}).  Despite the relatively long exposure times, 
these sources are rather faint for {\em INTEGRAL} with 20--50~keV ISGRI 
count rates between $0.18\pm 0.03$ and $0.35\pm 0.03$~s$^{-1}$ (compared 
to, e.g., a 20--60~keV count rate of 160 s$^{-1}$ for the Crab Pulsar
and Nebula).  Thus, the spectra of the IGR sources consist of a 
relatively small number of energy bins.  For IGR J16195--4945, 
IGR J16207--5129, and IGR J17195--4100, we are able to obtain 4 
energy bins in which the source is detected at a significance of 
at least 4-$\sigma$, and, for IGR J16167--4957, the source is only 
detected in 2 energy bins.

We fitted the ISGRI spectra with a power-law model, and the results
of these fits are given in Table~\ref{tab:integral}.  For 
IGR J16195--4945, IGR J16207--5129, and IGR J17195--4100, acceptable
fits are obtained with the power-law model with values of $\Gamma$
equal to $1.7\pm 0.8$, $1.9\pm 0.5$, and $2.8\pm 0.8$, respectively.
While, for IGR J16195--4945, the 90\% confidence error region for 
$\Gamma$ overlaps with the error region determined using the 
{\em Chandra} spectrum, the spectra for IGR J16207--5129 and 
IGR J17195--4100 are significantly steeper in the 20--50 keV
band than in the 0.3--10 keV {\em Chandra} band.  Although the 
spectrum for IGR J16167--4947 only has 2 energy bins, it is
still possible to calculate the value for $\Gamma$, and we
obtain $4.3^{+1.6}_{-1.4}$, which is significantly steeper
than the {\em Chandra} measurement.

The fact that the spectra are steeper in the ISGRI 20--50 keV
band when compared to the {\em Chandra} 0.3--10 keV band suggests
that there is a high energy cut off.  To investigate the cut off
and to compare the flux levels measured by the two instruments, 
we performed combined fits for the {\em Chandra} plus ISGRI spectra.  
As there is evidence that these sources are variable, and the 
{\em Chandra} and ISGRI spectra were not taken simultaneously, 
it is essential to leave the overall normalization between the
two instruments as a free parameter.  Leaving all parameters
free, we fitted the spectrum for each source with and without
an exponential cut off ({\tt highecut} in XSPEC).  The cut off 
is detected at the highest level of significance for IGR J16207--5129 
with an improvement in the fit from $\chi^{2}/\nu = 19.6/11$ to 
5.3/9, corresponding to an F-test significance of 99.7\%.  The 
cut off is also marginally statistically significant for 
IGR J16167--4957, with a 98.4\% chance that the cut off is required.  
The spectra for IGR J16195--4945 and IGR J17195--4100 allow for the 
possibility that there is a cut off in the ISGRI energy
range, but the cut off is not required.

While we allowed the $N_{\rm H}$, the power-law parameters, 
$\alpha$ in the pile-up model (for the three brighter sources), 
the cut off parameters ($E_{\rm cut}$ and $E_{\rm fold}$), and
the overall normalization between the two instruments to be 
free for the fits described in the previous paragraph, with
so many free parameters, the parameter values are very poorly
constrained.  Thus, we fixed the $N_{\rm H}$, power-law, and 
pile-up parameters to the values measured by {\em Chandra}, and 
left just the normalization between instruments and the exponential 
folding energy ($E_{\rm fold}$) as free parameters in the fit.  
We fixed the energy at which the exponential cut off begins 
($E_{\rm cut}$) to 10 keV, just above the {\em Chandra} bandpass.  
However, it should be noted that, while the {\em Chandra} spectra 
do not show any evidence for a spectral cut off, the quality of 
the spectra and the distortion of the spectra due to pile-up do 
not allow us to rule out the possibility that the cut off starts 
at a lower energy.  For IGR J16195--4945, IGR J16207--5129, 
IGR J16167--4957, and IGR J17195--4100, respectively, we obtained 
ISGRI relative normalizations of $1.1^{+0.9}_{-0.6}$, 
$0.23^{+0.10}_{-0.08}$, $1.4^{+2.5}_{-0.9}$, and $0.9^{+0.7}_{-0.4}$ 
as well as the following values for $E_{\rm fold}$:  $32^{+110}_{-14}$, 
$23^{+13}_{-6}$, $9^{+7}_{-3}$, and $20^{+18}_{-7}$ keV
(90\% confidence errors).  These values indicate that 
IGR J16207--5129 was relatively bright during the {\em Chandra} 
observation compared to the brightness detected by ISGRI in 
2003--2004, while the other sources were consistent with a
relative normalization of 1.0.  Also, it is notable that the
range of values we measure for $E_{\rm fold}$, 9--32 keV, are
similar to the values that have been measured for HMXB 
pulsars \citep{coburn02}.

\section{IR and Optical Identifications}

With the $0^{\prime\prime}.6$ {\em Chandra} positions for the 
4 IGR sources, we are able to search for counterparts to these 
sources at other wavelengths.  We searched the following catalogs: 
United States Naval Observatory (USNO-B1.0 and USNO-A2.0); 
Deep Near Infrared Survey of the Southern Sky (DENIS); 2 Micron 
All-Sky Survey (2MASS); and {\em Spitzer}'s Galactic Legacy 
Infrared Mid-Plane Survey Extraordinaire (GLIMPSE).  We also
performed optical and IR photometry at ESO's New Technology
Telescope, and the NTT observations and results are described
below.

The USNO catalogs are from optical surveys that cover most of the sky.  
For the USNO-B1.0, the position accuracies are $0^{\prime\prime}.2$ 
and the photometry is good to $\pm$0.3 magnitudes in $B$, $R$, and 
$I$.  USNO-A2.0 provides $0^{\prime\prime}.25$ positions and $B$ and 
$R$ measurements that are generally accurate to $\pm$0.5 magnitudes 
but may be somewhat worse that this in the southern hemisphere.
DENIS\footnote{See http://cdsweb.u-strasbg.fr/denis.html} is an optical 
and IR survey, concentrating on the southern hemisphere.  The 
position accuracies are $1^{\prime\prime}$, and accurate photometry 
is available in the $I$, $J$, and $K_{s}$ bands.  2MASS\footnote{See 
http://www.ipac.caltech.edu/2mass/releases/second/doc/} is an all-sky 
IR survey.  The catalog provides very good positions 
($0^{\prime\prime}.2$ accuracy) and accurate photometry in the $J$, 
$H$, and $K_{s}$ bands.  GLIMPSE uses the Infrared Array Camera (IRAC) 
on {\em Spitzer} to obtain short exposures of the Galactic plane, 
covering the Galactic longitude range between $10^{\circ}$ and 
$65^{\circ}$ on both sides of the Galactic center and Galactic 
latitudes within $1^{\circ}$ of the plane.  Images are obtained in 
four bands centered at wavelengths of 3.6, 4.5, 5.8, and 8.0 $\mu$m.  
The GLIMPSE team has compiled an on-line catalog\footnote{See
http://www.astro.wisc.edu/sirtf/} with lists of detected sources, 
including fluxes and positions accurate to $0^{\prime\prime}.3$.

\begin{figure}
\centerline{\includegraphics[width=0.50\textwidth]{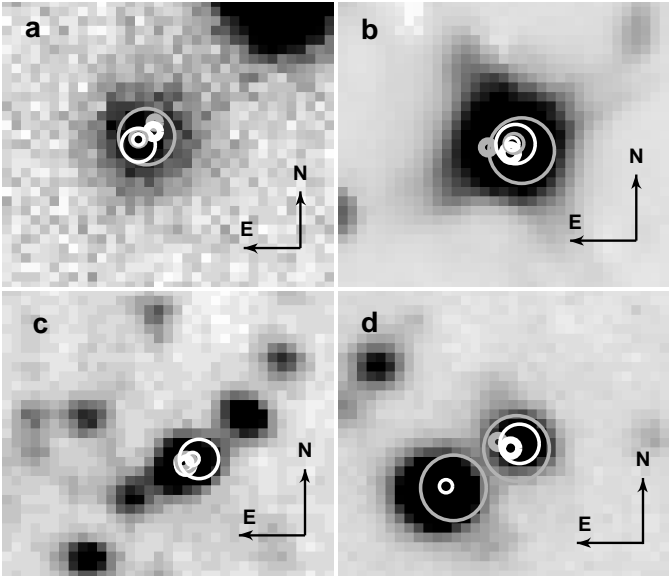}}
\caption{NTT images for (a) IGR J16195--4945 ($I$-band), 
(b) IGR J16207--5129 ($K_{s}$-band), (c) IGR J16167--4957 ($K_{s}$-band), 
and (d) IGR J17195--4100 ($K_{s}$-band).  The lengths of the N and E 
arrows are $2^{\prime\prime}$.  From smallest to largest, the error
circles are: 2MASS (thin white, $0^{\prime\prime}.2$ radius error circle),
USNO-B1.0 (thick grey, $0^{\prime\prime}.2$), USNO-A2.0 (thick white, 
$0^{\prime\prime}.25$), GLIMPSE (thin grey, $0^{\prime\prime}.3$), 
{\em Chandra} (thin white, $0^{\prime\prime}.6$), DENIS (thin grey, 
$1^{\prime\prime}$).\label{fig:ntt}}
\end{figure}

On 2004 July 10 between UT 0 and 1 hr, we obtained optical 
photometry in $B$, $V$, $R$, and $I$ bands of the fields of
IGR J16195--4945 and IGR J16207--5129 with the spectro-imager EMMI 
installed on the NTT.  We used the large field imaging of EMMI's 
detector, giving, after $2\times 2$ rebinning, a pixel size of 
$0^{\prime\prime}.333$ and a field of view of 
$9^{\prime}.9\times 9^{\prime}.0$.  We used an integration time of 
1~s per exposure. We observed 5 photometric standard stars from the 
optical standard star catalog of \cite{landolt92}: PG 1633+099, 
PG 1633+099A, PG 1633+099B, PG 1633+099C, and PG 1633+099D.  We 
also obtained IR photometry in $J$, $H$, and $K_{s}$ bands of 
IGR J16207--5129 (on 2004 July 8 at UT 4.67 hr), of IGR J16167--4957 
(on July 11 at UT 5.67 hr), and of IGR J17195--4100 (on July 11 at 
UT 6.83 hr) with the spectro-imager SofI installed on the NTT.  
We used the large field imaging of SofI's detector, giving a pixel 
size of $0^{\prime\prime}.288$ and a field of view of 
$4^{\prime}.94\times 4^{\prime}.94$.  These photometric observations 
were obtained by repeating a set of images for each filter with 9 
different $30^{\prime\prime}$ offset positions, including the targets, 
with an integration time of 60~s for each exposure, following the 
standard jitter procedure allowing for clean subtraction of the blank 
sky emission in IR.  On several occasions, we observed 3 photometric 
standard stars from the faint IR standard star catalog of 
\cite{persson98}: sj9157, sj9172, and sj9181. 
We used the Image Reduction and Analysis Facility (IRAF) suite to
perform data reduction, carrying out standard procedures of optical
and NIR image reduction, including flat-fielding and NIR sky subtraction.
We carried out aperture photometry, and we then transformed 
instrumental magnitudes into apparent magnitudes.  The targets
were close to the zenith during the observations.

Table~\ref{tab:oir} provides the results of our search of the catalogs, 
and also the results of our NTT observations.  For each of the 5 
catalogs, we list the closest source to the {\em Chandra} position and 
the measured magnitudes.  In Figure~\ref{fig:ntt}, we show an $I$-band 
NTT image for IGR J16195--4945, and $K_{s}$-band images for the other 
three sources.  The {\em Chandra} positions and the positions obtained 
from the catalogs are plotted on these images.

For IGR J16195--4945, the {\em Chandra} position lies on an apparently 
stellar IR source (based on an inspection of the 2MASS images) with a 
magnitude of $J = 13.57\pm 0.03$.  The 2MASS and GLIMPSE positions lie 
inside the {\em Chandra} error circle, indicating a good association 
between the X-ray and IR sources (see Figure~\ref{fig:ntt}a).  However, 
the USNO-A2.0 position lies on the edge of the {\em Chandra} error circle,
and the USNO-B1.0 position lies slightly outside.  This may indicate 
the presence of a second optical source blended with the counterpart.  
Due to the faintness of the source (or sources) it is not clear from 
the NTT $I$-band image whether there are two blended sources or not.  
However, we present analysis of the spectral energy distributions
(SEDs) below, and the IGR J16195--4945 SED (see 
Figure~\ref{fig:broadband}a) supports the possibility that there are 
two blended sources.

For IGR J16207--5129, the {\em Chandra} position lies on a bright 
IR source with $J = 10.44\pm 0.02$.  This source is present in all 5 
catalogs, and all the positions are consistent with the {\em Chandra} 
source except for the USNO-B1.0 position, which lies just outside the 
{\em Chandra} error circle (see Figure~\ref{fig:ntt}b).  While this 
could indicate the presence of an interloper, the SED does not suggest 
any contamination by other sources (see Figure~\ref{fig:broadband}b).  

For IGR J16167--4957, the {\em Chandra} position lies on a source with 
$J = 14.86\pm 0.06$.  The source appears in the GLIMPSE, 2MASS, and USNO 
catalogs, and all positions lie within the {\em Chandra} error circle 
(see Figure~\ref{fig:ntt}c).  Although the 2MASS images suggest
that the source is extended to the South-East, the NTT $K_{s}$-band 
image shows that there is actually a blend of at least 3 sources.  The 
IGR J16167--4957 {\em Chandra} position is consistent with the brightest 
of the blended IR sources, but there is some suggestion in the NTT image 
that the brightest ``source" may also be a blend.

For IGR J17195--4100, the {\em Chandra} position is near two sources 
that are clearly blended in the 2MASS images.  The position of the 2MASS 
source is $2^{\prime\prime}.6$ away from the {\em Chandra} position, 
indicating that the bright IR source is not IGR J17195--4100.  With 
$I$-band and IR coverage, both sources appear in the DENIS survey, and 
the optically-brighter, North-West source is the one that is consistent 
with the {\em Chandra} position.  Our results confirm the association 
between IGR J17195--4100 and USNO-B1.0 0489--0511283.

\section{Spectral Energy Distributions}

\begin{figure}
\centerline{\includegraphics[width=0.50\textwidth]{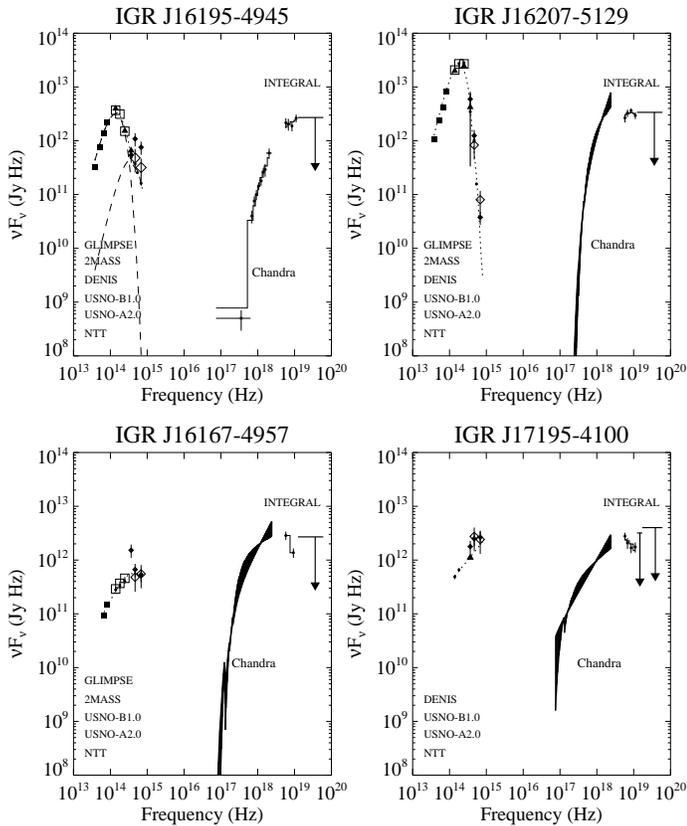}}
\caption{\footnotesize Spectral energy distributions (SEDs) for 
the four IGR sources.  The IR and optical fluxes come from the 
measurements given in Table~\ref{tab:oir}, and this part of the 
spectrum has been fitted with a blackbody model modified by the 
effects of extinction.  The dotted line indicates the best fit 
(see text).  {\em Chandra} and {\em INTEGRAL} measurements are 
also shown.  For IGR J16207--5129, IGR J16167--4957, and 
IGR J17195--4100, our best estimate of the spectrum in the 
{\em Chandra} energy range is the model (solid line).  For the
optical/IR data, the different symbols are: 
GLIMPSE (filled squares), 2MASS (open squares), DENIS (filled 
triangles), USNO-B1.0 (filled diamonds), USNO-A2.0 (open 
diamonds), NTT (filled circles).\label{fig:broadband}}
\end{figure}

Figure~\ref{fig:broadband} shows the spectral energy distributions 
(SEDs) for the four sources.  These include the optical and IR 
measurements given in Table~\ref{tab:oir}, the spectrum in the 
{\em Chandra} bandpass, and {\em INTEGRAL} measurements.  It
should be noted that the optical and IR fluxes are not dereddened, 
and absorption is not removed from the {\em Chandra} spectra.  The 
SED measurements are spread over a period of years, so discontinuities 
in the SED may be due to source variability.  For the IR and optical 
parts of the SED, a check on variability is provided by the fact that
we have multiple measurements for many of the photometric bands.  A
comparison shows that the IR fluxes are quite stable for 
IGR J16195--4945, IGR J16207--5129, and IGR J16167--4957.  For
IGR J16195--4945, the 1-$\sigma$ error regions overlap for the
$J$- and $K_{s}$-band measurements from DENIS and 2MASS.  For
IGR J16207--5129, the DENIS, 2MASS, and NTT $J$- and $K_{s}$-band
measurements only range from 10.38--10.54 and 9.13--9.18, respectively.
For IGR J16167--4957, the 2MASS and NTT measurements are consistent
at the 1-$\sigma$ level for $J$, $H$, and $K_{s}$.  For 
IGR J16207--5129, the optical flux also appears to be rather stable, 
but for IGR J16195--4945 and IGR J16167--4957, the agreement between 
the various optical measurements is not as good (see 
Figure~\ref{fig:broadband}).  For IGR J17195--4100, the agreement in 
the optical bands is relatively good, but this source has the sparsest 
optical/IR measurements.

For IGR J16195--4945, we simply show in Figure~\ref{fig:broadband}a
the data and the best fit
model for {\em Chandra}; however, for the other three sources, 
the spectrum is significantly distorted by pile-up, and we show
the range of flux measurements based on the error ranges for the
model parameters.  For IGR J16207--5129, the flux range shown 
reflects the 90\% confidence error range on $\Gamma$, which is
0.0--1.1.  For IGR J16167--4957 and IGR J17195--4100, the 
$\Gamma$ ranges are 0.7--1.6 and 0.8--1.4, respectively.  The
{\em INTEGRAL} 20--50 keV measurements and 50--250 keV upper 
limits are shown.  As described in \S$5$, the fits shown include
a high energy cut off and a free normalization between {\em Chandra}
and ISGRI.  IGR J16207--5129 is the only source for which the
normalization for ISGRI relative to {\em Chandra} is required to
be less than 1.0, and this is apparent in Figure~\ref{fig:broadband}.

To characterize the shape of the optical and IR continuum, we 
fitted each of the SEDs with four models:  a single blackbody,
a single power-law, the sum of 2 blackbodies, and a blackbody
plus a power-law.  Presumably, blackbody components could be
indicative of thermal components from a star, but we consider 
the power-law to be a phenomenological component.  The models 
also account for interstellar extinction using the analytical 
approximation of \cite{ccm89}.  This adds a single free parameter, 
$A_{V}$, to the single component models, while, for the
two-component models, we allow the components to have different
extinctions.  We used $\chi^{2}$-minimization when fitting the SEDs, 
and, for IGR J16195--4945 and IGR J16207--5129, all of the models 
resulted in formally very poor fits with the best models leading to 
$\chi^{2}/\nu = 128/14$ and $\chi^{2}/\nu = 684/16$ for the two 
sources, respectively.  Such poor fits may indicate inadequacies 
of our models, but they are also at least in part due to systematic 
errors, which may be related to the variability discussed above 
or possibly due to, e.g., uncertainties in magnitude/flux conversion.  
While it is clear that systematic errors must be included for these 
2 sources, it is not clear which data points are the most suspect.  
Thus, we included systematic errors by simply multiplying the 
existing errors on all of the optical and IR data points by a 
constant factor.  We adjusted the constant factor until we achieved
a reduced-$\chi^{2}$ of 1.0 for at least one of the models.  For 
IGR J16195--4945 and IGR J16207--5129, the constant factors are 
2.2 and 6.5, respectively.  For the other 2 sources, 
IGR J16167--4957 and IGR J17195--4100, we did not need to include 
any systematic errors.  The SED fit results are presented in 
Table~\ref{tab:seds}, and here, we describe the results for each 
source in turn.  

For IGR J16195--4945, neither of the single component models
provide an acceptable fit to the optical/IR SED, and a two-component 
model is required at high significance.  The two blackbody and
blackbody plus power-law models provide acceptable fits, and both
imply that the IR flux is dominated by a high temperature star 
($>$9400 K) with high extinction ($A_{V}\sim 18$).  Also, both 
models imply that the component that dominates the optical flux 
has a significantly lower extinction ($A_{V} < 7.1$).  This SED 
and the shift between the optical and IR source positions reported 
in \S$6$ provide evidence that the optical component comes from 
an interloping star, while the IR component likely comes from 
IGR J16195--4945.  

Our highest quality optical/IR SED is the one for IGR J16207--5129.
For this SED, two-component models do not provide any improvement
over the single component models, and the blackbody and power-law
models provide fits of roughly the same quality.  The power-law 
index is $\alpha = -2.29^{+0.24}_{-0.28}$, which is nearly consistent
with the value of $\alpha = -2$ expected for a blackbody.  The
fits strongly suggest that the spectrum is thermal as would be
expected for stellar emission.  From the blackbody fit, the
value obtained for the extinction, $A_{V} = 10.8^{+0.3}_{-0.8}$
is very close to the value of $A_{V} = 9.5$ obtained by converting
the Galactic column density of $N_{\rm H} = 1.7\times 10^{22}$ cm$^{-2}$
to optical extinction using the \cite{ps95} relationship.

A blackbody model does not provide a good fit to the IGR J16167--4957 
and IGR J17195--4100 SEDs with values of $\chi^{2}/\nu$ of 54/10 and 
26/6, respectively.  However, both SEDs are adequately described by 
a power-law model with indices of $\alpha = -0.36^{+0.17}_{-0.18}$ and 
$\alpha = 0.17^{+0.12}_{-0.47}$, respectively.  The power-law fits
for both sources imply low extinction, which is consistent with the 
values of $N_{\rm H}$ measured via fits to the X-ray spectrum.
Although we fitted the SEDs for these two sources with the
two component models, the improvements in $\chi^{2}$ do not
justify the addition of another component, and the parameters
for the two-component fits are very poorly constrained.

\section{Discussion}

In the following, we discuss the detailed constraints on the nature of 
each source, considering the information we have obtained from 
{\em Chandra}, {\em INTEGRAL}, and the optical/IR SEDs.

\subsection{IGR J16195--4945}

Of our 4 sources, IGR J16195--4945 is the most similar to the class
of obscured IGR sources.  At $N_{\rm H} = (7^{+5}_{-3})\times 10^{22}$
cm$^{-2}$, the X-ray measured column density is higher than the Galactic 
value of $2.2\times 10^{22}$~cm$^{-2}$, suggesting the possibility that
the X-ray source is intrinsically absorbed.  Due to the possibility of 
intrinsic absorption, we cannot use the X-ray-measured $N_{\rm H}$ to 
constrain the distance to the source; however, because the optical and 
IR emission from the system is likely to predominantly come from regions 
that are not as affected by intrinsic absorption, such as the companion 
star, we expect that the optical extinction is dominated by interstellar 
material.  Thus, the fact that $A_{V} = 17.5^{+0.8}_{-2.3}$ for the 
extinction on the blackbody component is as large or larger than the
value obtained by converting the Galactic $N_{\rm H}$ to optical
extinction ($A_{V} = 12.3$), is likely to indicate that we are looking 
at the source through much of the Galaxy. 
The distance to the Norma-Cygnus arm along this line of sight is
$\sim$5 kpc \citep{russeil03}, and, although we cannot formally constrain
the distance to the source, we assume a fiducial value of 5~kpc.
From the unabsorbed fluxes reported in Table~\ref{tab:spectra} and
\ref{tab:integral}, the implied X-ray luminosities are 
$1.4\times 10^{34}$ ($d$/5 kpc)$^{2}$ (0.3--10 keV) ergs~s$^{-1}$ and 
$5.8\times 10^{34}$ ($d$/5 kpc)$^{2}$ ergs~s$^{-1}$ (20--50 keV).  

\begin{figure}
\centerline{\includegraphics[width=0.50\textwidth]{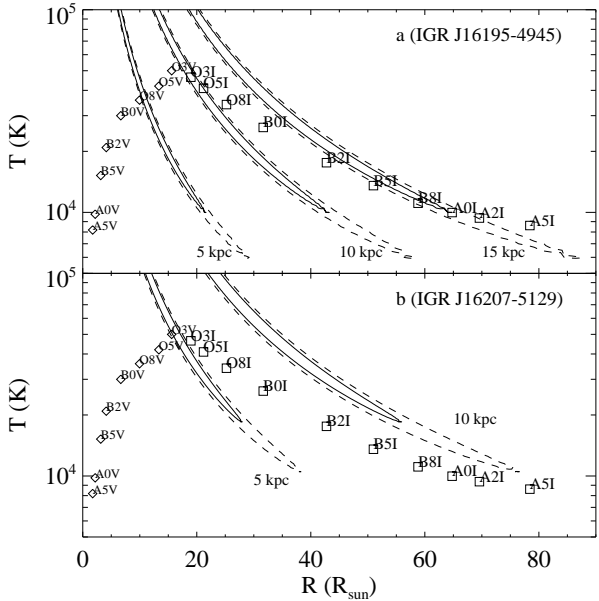}}
\caption{Error contours for IGR J16195--4945 ({\em a}) and 
IGR J16207--5129 ({\em b}) for the blackbody parameters in the
optical/IR SEDs.  Contours are shown for the stellar radius ($R$) 
and temperature ($T$), assuming various distances.  The solid 
contours are 68\% confidence ($\Delta\chi^{2} = 2.3$), and the 
dashed contours are 90\% confidence ($\Delta\chi^{2} = 4.6$).  
Radii and temperatures for various spectral types from 
\cite{djn87} are overplotted, allowing us to constrain the 
likely spectral types of the companions in these systems.
\label{fig:contour2}}
\end{figure}

The IR portion of the SED is consistent with a single-temperature
blackbody, suggesting that it is more likely that the IR emission
comes from a stellar companion than from, e.g., an accretion disk.
The lack of variability in the IR emission discussed in \S$7$
also provides evidence that the emission comes from a star.  
The fit to the SED gives a lower limit on the blackbody temperature
of $T > 9400$~K, and this indicates a spectral type earlier than 
A1--A2 regardless of whether the star is a main sequence star or a
supergiant \citep{djn87,cox00}.  The question of whether the companion 
is a main sequence star or a supergiant depends sensitively on the 
distance to the source.  In the fit to the SED, the parameters
$T$ and $R/d$ have a significant level of degeneracy, and, in 
Figure~\ref{fig:contour2}a, we show 68\% and 90\% confidence error 
contours for $T$ and $R$ for assumed distances of 5~kpc, 10~kpc, 
and 15~kpc along with temperatures and radii for stars with various 
spectral types.  If the distance is $\sim$3--9 kpc, the companion 
is a main sequence B or O-type star.  If the distance is 
$\sim$9--15 kpc, a large range of supergiant A, B, and O-type 
spectral types are possible.

\subsection{IGR J16207--5129}

Like IGR J16195--4945, the IGR~J16207--5129 $N_{\rm H}$ is 
somewhat (about a factor of 2) higher than the Galactic value, 
allowing for, but not requiring, some intrinsic absorption.
From the optical/IR SED, $A_{V} = 10.8^{+0.3}_{-0.8}$, and this
value suggests a relatively large distance for this source.  
Thus, it is reasonable to take 5~kpc, the distance to the 
Norma-Cygnus arm along the line of sight, as a fiducial distance.  
The implied 0.3--10 keV luminosity is $1.3\times 10^{35}$ 
($d$/5 kpc)$^2$ ergs~s$^{-1}$, about an order of magnitude higher 
than IGR J16195--4945, and the 20--50 keV luminosity is 
$8.3\times 10^{34}$ ($d$/5 kpc)$^2$ ergs~s$^{-1}$.  

In this case, the optical/IR SED consists of a single-temperature
blackbody, and it is very likely that this emission comes from the
stellar companion.  The lower limit on the stellar temperature
is $>$18,000 K, indicating that this system harbors a very hot, and
likely massive, star.  Figure~\ref{fig:contour2} shows the error 
contours for the parameters $T$ and $R$ assuming distances of 
5 and 10~kpc.  At $\sim$3--5 kpc, the companion is a main 
sequence B or O-type star.  At $\sim$5--9 kpc, various supergiant
types are possible.  At 10~kpc, the range of possible temperatures and
radii are not consistent with any known spectral type, indicating
that this system is closer than 10~kpc.

\subsection{IGR J16167--4957}

Although this source is close to the Galactic plane at $b = 
+0.50^{\circ}$ and is within $0.5^{\circ}$ of IGR J16195--4945
and $1.7^{\circ}$ of IGR J16207--5129, its X-ray spectrum indicates 
a column density of $N_{\rm H} = (5^{+3}_{-2})\times 10^{21}$ 
cm$^{-2}$, which is an order of magnitude lower than the value for 
IGR J16195--4945 and is also lower than the Galactic column density 
of $2.2\times 10^{22}$ cm$^{-2}$ along this line of sight.  The 
measured $N_{\rm H}$ corresponds to an extinction of $A_{V} = 
2.8^{+1.7}_{-1.1}$, and, although the optical/IR SED is fit with 
a phenomenological power-law model, the implied extinction is even 
lower, $A_{V} = 1.3^{+0.6}_{-0.5}$.  While the inhomogeneous 
distribution of gas and dust in the Galaxy make it difficult to use 
extinction measurements to determine source distances in any rigorous 
way, here, we make an estimate using the values from the \cite{ds94} 
study, which includes UV extinction measurements of nearly 400 
early-type stars through various lines of sight through the Galaxy 
and with distances as large as 10 kpc.  In that study, the average
extinction per kpc is $A_{V}/d = 0.45$ magnitudes/kpc.  For
IGR J16167--4957, the extinction inferred from the X-ray spectrum
allows the distance to be anywhere from 2 to 10 kpc, but the optical/IR
extinction implies a distance of $2.9^{+1.3}_{-1.1}$ kpc.  This 
estimate is marginally consistent with the source being in the
Norma-Cygnus arm at $\sim$5 kpc, but the line of sight also
crosses the Scutum-Crux arm at $\sim$3 kpc.  Given the lower
extinction for this source compared to IGR J16195--4945 and
IGR J16207--5129, which are likely in the Norma-Cygnus arm, we
take 3 kpc as a fiducial distance for IGR J16167--4957, and
the X-ray luminosities are $4.7\times 10^{34}$ ($d$/3 kpc)$^{2}$ 
ergs~s$^{-1}$ (0.3--10 keV) and $1.7\times 10^{34}$ ($d$/3 kpc)$^{2}$
ergs~s$^{-1}$ (20--50 keV).  

The interpretation for the shape of the optical/IR SED is not
immediately clear, but the fact that it is not consistent with
a single temperature blackbody suggests that the emission is not
simply from a companion star.  The power-law index of 
$-0.36^{+0.17}_{-0.18}$ is considerably flatter than a blackbody, 
and this could indicate a multi-temperature blackbody as might
be expected from an accretion disk.  Regardless of the dominant 
contributor to the optical/IR emission, we can take the measured 
magnitudes as an upper limit on the contribution from a putative 
optical companion.  If we assume a distance and an extinction at 
the upper ends of the derived ranges, 4.2 kpc and $A_{V} = 2.8$, 
respectively, and interpolate between the measured $B$ and $R$
magnitudes to infer a $V$-band magnitude of 16.5, the limit on 
the absolute magnitude is $M_{V} > 0.6$.  Unless the distance is 
very much larger than 4.2 kpc, this rules out a supergiant companion, 
and the companion's spectral type is later than A0V.  Thus, 
IGR J16167--4957 is not an HMXB.

\subsection{IGR J17195--4100}

At $l = 346.98^{\circ}$ and $b = -2.14^{\circ}$, this source
is $\sim$$14^{\circ}$ from the other 3 sources, and is slightly
off the Galactic plane.  The measured upper limit on the column 
density of $N_{\rm H} < 2\times 10^{21}$ cm$^{-2}$ is considerably 
lower than the Galactic value of $7.7\times 10^{21}$ cm$^{-2}$, 
indicating that the distance to the source may be relatively low.  
The $N_{\rm H}$ from the X-ray spectrum implies $A_{V} < 1.18$, 
and the optical/IR SED gives an upper limit of $A_{V} < 0.96$.  
Using the average value of $A_{V}/d$ from \cite{ds94} indicates 
a distance upper limit in the range $d < 2.1$--2.6 kpc.  The 
upper limit on the distance of 2.6 kpc implies upper limits on 
the X-ray luminosities of $<$$2.0\times 10^{34}$ ergs~s$^{-1}$ 
(0.3--10 keV) and $<$$1.6\times 10^{34}$ ergs~s$^{-1}$ (20--50 keV). 

Like IGR J16167--4957, the physical interpretation for the power-law 
shape of the optical/IR SED is not clear.  We can take the optical/IR 
magnitudes observed as upper limits on the contribution from a 
companion star.  Interpolating between the $B$ and $R$-bands, we 
estimate that $V = 14.8$.  This, along with the highest upper 
limits we measure for the extinction, $A_{V} < 1.18$, and distance, 
$d < 2.6$ kpc, imply an upper limit on the absolute magnitude
of $M_{V} < 1.5$.  This rules out a supergiant companion or an 
OB main sequence star.  The earliest spectral type possible is
$\sim$A3V.

\section{Summary of the Nature of the 4 IGR Sources}

Table~\ref{tab:summary} contains a summary of the results described
in \S$8$, including constraints on the source distance, X-ray
luminosities, and possible spectral types.  Our basic conclusion
is that IGR J16195--4945 and IGR J16207--5129 are very likely
HMXBs and are at distances consistent with the Norma-Cygnus arm
(although the distances are not well enough constrained to 
say definitively that they are in this spiral arm), while 
the other two IGR sources (IGR J16167--4957 and IGR J17195--4100) 
are nearer and are not HMXBs.  Assuming distances of 5~kpc for
the 2 HMXBs, the 0.3--10 keV luminosities are $10^{34-35}$
ergs~s$^{-1}$, which is, not surprisingly, significantly lower 
than the brightest HMXB pulsars (e.g., Vela X-1, Cen X-3), but 
the luminosities are not atypical when considering the full range 
of HMXB pulsar luminosities.  

Of the 4 IGR sources, only IGR J16195--4945 has a column density that 
is consistent with the source having significant intrinsic absorption.  
While IGR J16207--5129 could conceivably also have some level of 
intrinsic absorption, comparing its column density to the Galactic 
value suggests that the intrinsic $N_{\rm H}$ is not more than 
$\sim$$1\times 10^{22}$ cm$^{-2}$ while many IGR sources have 
$N_{\rm H} \sim 10^{23-24}$ cm$^{-2}$.  From the perspective of IGR 
sources and HMXB evolution \citep{dean05}, the very high column density 
IGR sources may be in an evolutionary stage during which the neutron star 
is spiraling toward its companion, becoming embedded in the wind from the 
high-mass companion.  In this picture, we may be seeing IGR J16207--5129 
early in this evolutionary phase so that the neutron star is still in a 
low density portion of the wind.  Source-to-source differences in $N_{\rm H}$ 
could also be indicative of variations in the strength of the stellar 
wind over time.  

After submission of this work, two other reports provided further 
information about the HMXBs.  Using {\em INTEGRAL}, \cite{sguera06} 
found a 1.5 hour X-ray outburst from IGR J16195--4945, indicating
that this source may belong to the class of supergiant fast X-ray
transients.  Also, for IGR J16207--5129, \cite{masetti06_atel} report on
optical spectra of USNO-A2.0 0375-27093111 (i.e., the same star
that we have determined to be the counterpart), and find that the 
spectrum includes H$\alpha$ in emission and is consistent with that 
of an HMXB.

\cite{masetti06_atel} also report on the optical spectra of the stars 
that we have (independently) determined to be the optical counterparts
of IGR J17195--4100 and IGR J16167--4957.  These stars show Hydrogen
Balmer series emission lines as well as HeI and HeII emission lines.
While this indicates that they are either Cataclysmic Variables (CVs)
or LMXBs, the hard X-ray spectra would be very unusual for LMXBs
\citep[e.g.,][]{muno04}.  A CV interpretation is more likely, and 
\cite{masetti06_atel} also come to this conclusion.

\acknowledgments

JAT acknowledges partial support from {\em Chandra} award number
GO5-6037X issued by the {\em Chandra X-Ray Observatory Center}, which
is operated by the Smithsonian Astrophysical Observatory for and on 
behalf of the National Aeronautics and Space Administration (NASA), 
under contract NAS8-03060.  JAT acknowledges partial support from a 
NASA {\em INTEGRAL} Guest Observer grant NNG05GC49G.  LF acknowledges 
partial funding from the Italian Space Agency (ASI) under contract 
I/R/046/04 and from MIUR under contract COFIN 2004-023189. We thank 
the referee for helpful comments that improved this manuscript.
This work is based, in part on ESO observations through program
\#073.D-0339.  We thank the referee for helpful comments that 
improved this manuscript.  This publication makes use of data products 
from the Two Micron All Sky Survey, which is a joint project of the 
University of Massachusetts and the Infrared Processing and Analysis 
Center/California Institute of Technology, funded by NASA and the 
National Science Foundation.  This research has made use of the USNOFS 
Image and Catalogue Archive operated by the United States Naval 
Observatory, Flagstaff Station as well as the SIMBAD database, operated 
at CDS, Strasbourg, France.  We have also used data from {\em Spitzer}'s 
Galactic Legacy Infrared Mid-Plane Survey Extraordinaire (GLIMPSE) as 
well as the Deep Near Infrared Survey of the Southern Sky (DENIS).



\begin{table}
\caption{{\em Chandra} Observations\label{tab:obs}}
\begin{minipage}{\linewidth}
\footnotesize
\begin{tabular}{cccccc} \hline \hline
Obs ID & Target & $l$\footnote{Galactic longitude in degrees.} & $b$\footnote{Galactic latitude in degrees.} & Start Time & Exposure Time (s)\\ \hline \hline
5471 & IGR J16195--4945 & 333.56 & +0.34 & 2005 April 29, UT 17:25 & 4,752\\
5472 & IGR J16207--5129 & 332.46 & --1.05 & 2005 June 25, UT 03:17 & 5,109\\
5473 & IGR J16167--4957 & 333.06 & +0.50 & 2005 June 13, UT 19:14 & 4,979\\
5474 & IGR J17195--4100 & 346.98 & --2.14 & 2005 July 25, UT 12:57 & 4,701\\ \hline
\end{tabular}
\end{minipage}
\end{table}

\begin{table}
\caption{{\em Chandra} Count Rates and Positions\label{tab:loc}}
\begin{minipage}{\linewidth}
\footnotesize
\begin{tabular}{cccccc} \hline \hline
IGR Name & CXOU Name & ACIS Rate\footnote{Count rate detected by the ACIS-I instrument in the 0.3--10 keV bandpass.} & R.A.\footnote{Right Ascension (equinox J2000).  The radius of the error circle is $0^{\prime\prime}\!.6$.} & Decl.\footnote{Declination (equinox J2000).  The radius of the error circle is $0^{\prime\prime}\!.6$.} & X-Ray Identification\\ \hline \hline
J16195--4945 & J161932.2--494430 & 0.039 & $16^{\rm h}19^{\rm m}32^{\rm s}\!.20$ & $-49^{\circ}44^{\prime}30^{\prime\prime}\!.7$ & AX~J161929--4945(?)\\
J16207--5129 & J162046.2--513006 & 0.13 & $16^{\rm h}20^{\rm m}46^{\rm s}\!.26$ & $-51^{\circ}30^{\prime}06^{\prime\prime}\!.0$ & --\\
J16167--4957 & J161637.7--495844 & 0.18 & $16^{\rm h}16^{\rm m}37^{\rm s}\!.74$ & $-49^{\circ}58^{\prime}44^{\prime\prime}\!.5$ & 1RXS~J161637.2--495847\\
J17195--4100 & J171935.8--410053 & 0.15 & $17^{\rm h}19^{\rm m}35^{\rm s}\!.88$ & $-41^{\circ}00^{\prime}53^{\prime\prime}\!.6$ & 1RXS~J171935.6--410054\\ \hline
\end{tabular}
\end{minipage}
\end{table}

\begin{table}
\caption{{\em Chandra} Spectral Results\label{tab:spectra}}
\begin{minipage}{\linewidth}
\footnotesize
\begin{tabular}{ccccccc} \hline \hline
IGR Name & $N_{\rm H}$\footnote{Errors in this table are at the 90\% confidence level ($\Delta$$C = 2.7$).} & $\Gamma$ & Flux\footnote{Unabsorbed 0.3--10 keV flux in units of $10^{-12}$ erg~cm$^{-2}$~s$^{-1}$.} & $\alpha$\footnote{The grade migration parameter in the pile-up model \citep{davis01}.  The probability that $n$ events will be piled together but will still be retained after data filtering is $\alpha^{n-1}$.} & Fit Statistic\footnote{The Cash statistic for the best fit model.  In each case, the spectra include 663 energy bins.} & Galactic $N_{\rm H}$\footnote{The column density through the Galaxy from \cite{dl90}.}\\
 & ($\times 10^{22}$ cm$^{-2}$) & & & & & ($\times 10^{22}$ cm$^{-2}$)\\ \hline \hline
J16195--4945 & $7^{+5}_{-3}$ & $0.5^{+0.9}_{-0.7}$ & $4.6^{+2.1}_{-0.8}$ & -- & 453 & 2.2\\
J16207--5129 & $3.7^{+1.4}_{-1.2}$ & $0.5^{+0.6}_{-0.5}$ & $42^{+9}_{-7}$ & $0.43^{+0.10}_{-0.09}$ & 667 & 1.7\\
J16167--4957 & $0.5^{+0.3}_{-0.2}$ & $1.1^{+0.5}_{-0.4}$ & $44^{+11}_{-10}$ & $0.62^{+0.06}_{-0.07}$ & 701 & 2.2\\ 
J17195--4100 & $0.08^{+0.13}_{-0.08}$ & $1.1\pm 0.3$ & $25^{+9}_{-4}$ & $0.64^{+0.09}_{-0.10}$ & 743 & 0.77\\ \hline
\end{tabular}
\end{minipage}
\end{table}

\begin{table}
\caption{{\em INTEGRAL} Spectral Results\label{tab:integral}}
\begin{minipage}{\linewidth}
\footnotesize
\begin{tabular}{cccccc} \hline \hline
IGR Name & Exposure Time (ks) & Count Rate\footnote{20--50 keV ISGRI count rate.} & $\Gamma$ & Flux\footnote{20--50 keV flux in units of $10^{-12}$ erg~cm$^{-2}$~s$^{-1}$.} & $\chi^{2}/\nu$\\ \hline \hline
J16195--4945 & 167 & $0.24\pm 0.03$ & $1.7\pm 0.8$ & $19\pm 3$ & 1.3/2\\
J16207--5129 & 165 & $0.35\pm 0.03$ & $1.9\pm 0.5$ & $28\pm 3$ & 2.5/2\\
J16167--4957 & 167 & $0.18\pm 0.03$ & $4.3^{+1.6}_{-1.4}$ & $16^{+3}_{-4}$ & 0/0\\
J17195--4100 & 495 & $0.23\pm 0.02$ & $2.8\pm 0.8$ & $19\pm 3$ & 1.1/2\\ \hline
\end{tabular}
\end{minipage}
\end{table}

\begin{table}
\caption{Optical and IR magnitudes or fluxes\label{tab:oir}}
\begin{minipage}{\linewidth}
\footnotesize
\begin{tabular}{ccccc} \hline \hline
  & IGR J16195--4945 & IGR J16207--5129 & IGR J16167--4957 & IGR J17195--4100\\ \hline \hline
\multicolumn{5}{c}{USNO-B1.0} \\
\hline
Name\footnote{Name of the nearest star in the catalog.} & 0402-0529810 & 0384-0560875 & 0400-0527262 & 0489-0511283\\
Separation\footnote{Separation between the {\em Chandra} position and the nearest star in the catalog.} & $1^{\prime\prime}.03$ & $0^{\prime\prime}.81$ & $0^{\prime\prime}.53$ & $0^{\prime\prime}.70$\\
$B$ & $16.5\pm 0.3$ & $19.7\pm 0.3$ & $16.8\pm 0.3$ & $15.1\pm 0.3$\\
$R$ & $15.3\pm 0.3$ & $15.2\pm 0.3$ & $15.8\pm 0.3$ & $14.4\pm 0.3$\\
$I$ & $15.6\pm 0.3$ & $13.0\pm 0.3$ & $14.5\pm 0.3$ & $14.3\pm 0.3$\\ \hline
\multicolumn{5}{c}{USNO-A2.0} \\
\hline
Name$^{a}$ & 0375-27014824 & 0375-27093111 & 0375-26829054 & 0450-27095307\\
Separation$^{b}$ & $0^{\prime\prime}.72$ & $0^{\prime\prime}.31$ & $0^{\prime\prime}.45$ & $0^{\prime\prime}.31$\\
$B$ & $17.4\pm 0.5$ & $18.9\pm 0.5$ & $16.8\pm 0.5$ & $15.2\pm 0.5$\\
$R$ & $16.2\pm 0.5$ & $15.6\pm 0.5$ & $16.2\pm 0.5$ & $14.3\pm 0.5$\\ \hline
\multicolumn{5}{c}{DENIS} \\
\hline
Name$^{a}$ & J161932.1--494430 & J162046.2--513006 & --- & J171935.8--410053\\
Separation$^{b}$ & $0^{\prime\prime}.42$ & $0^{\prime\prime}.29$ & --- & $0^{\prime\prime}.10$\\
$I$ & $15.38\pm 0.05$ & $13.4\pm 1.0$ & --- & $14.81\pm 0.04$\\
$J$ & $13.55\pm 0.08$ & $10.54\pm 0.05$ & --- & ---\\
$K_{s}$ & $10.92\pm 0.06$ & $9.17\pm 0.06$ & --- & ---\\ \hline
\multicolumn{5}{c}{2MASS} \\
\hline
Name$^{a}$ & J16193220--4944305 & J16204627--5130060 & J16163776--4958445 & J17193608--4100548\\
Separation$^{b}$ & $0^{\prime\prime}.14$ & $0^{\prime\prime}.11$ & $0^{\prime\prime}.26$ & $2^{\prime\prime}.6$\\
$J$ & $13.57\pm 0.03$ & $10.44\pm 0.02$ & $14.86\pm 0.06$ & ---\footnote{This 2MASS source is too far from the {\em Chandra} position to be associated.}\\
$H$ & $11.96\pm 0.03$ & $9.62\pm 0.02$ & $14.28\pm 0.09$ & ---$^{c}$\\
$K_{s}$ & $11.00\pm 0.02$ & $9.13\pm 0.02$ & $13.76\pm 0.10$ & ---$^{c}$\\ \hline
\multicolumn{5}{c}{GLIMPSE} \\ 
\hline
Name$^{a}$ & G333.5571+00.3390 & G332.4590--01.0501 & G333.0560+00.4962 & ---\\
Separation$^{b}$ & $0^{\prime\prime}.21$ & $0^{\prime\prime}.07$ & $0^{\prime\prime}.34$ & ---\\
3.6 $\mu$m & $26.5\pm 1.1$ mJy & $100\pm 3$ mJy & $1.79\pm 0.16$ mJy & ---\\
4.5 $\mu$m & $20.3\pm 1.1$ mJy & $61\pm 3$ mJy & $1.40\pm 0.16$ mJy & ---\\
5.8 $\mu$m & $14.3\pm 1.0$ mJy & $46\pm 2$ mJy & --- & ---\\
8.0 $\mu$m & $8.5\pm 0.5$ mJy & $28.4\pm 0.9$ mJy & --- & ---\\ \hline
\multicolumn{5}{c}{New Technology Telescope}\\
\hline
$B$ & $18.14\pm 0.06$ & $19.8\pm 0.1$   & ---           & ---\\
$V$ & $17.22\pm 0.05$ & $17.74\pm 0.06$ & ---           & ---\\
$R$ & $16.42\pm 0.05$ & $15.38\pm 0.03$ & ---           & ---\\
$I$ & $15.54\pm 0.03$ & $13.58\pm 0.02$ & ---           & ---\\
$J$ & ---             & $10.38\pm 0.02$ & $15.0\pm 0.1$ & $14.1\pm 0.1$\\
$H$ & ---             & $9.60\pm 0.02$  & $14.4\pm 0.1$ & $13.65\pm 0.07$\\
$K_{s}$ & ---         & $9.18\pm 0.04$  & $13.8\pm 0.1$ & $13.2\pm 0.1$\\ \hline
\end{tabular}
\end{minipage}
\end{table}

\begin{table}
\caption{IR/Optical SED Fits\label{tab:seds}}
\begin{minipage}{\linewidth}
\footnotesize
\begin{tabular}{cccccc} \hline \hline
  & IGR J16195--4945 & IGR J16207--5129 & IGR J16167--4957 & IGR J17195--4100\\ \hline \hline
\multicolumn{5}{c}{Blackbody} \\ 
\hline
$A_{V}$ & $<$0.16 & $10.8^{+0.3}_{-0.8}$ & $<$1.16 & $<$2.84\\
$T$ (K) & $1944^{+39}_{-42}$ & $>$18000 & $2790^{+350}_{-70}$ & $4100^{+5500}_{-200}$\\
$R/d$ (\Rsun/kpc)\footnote{$R/d$ is the ratio of the radius of the spherical blackbody (in \Rsun) to the distance to the source (in kpc).} & $10.3\pm 0.6$ & $<$5.53 & $1.84\pm 0.21$ & $1.40^{+0.14}_{-0.62}$\\
$\chi^{2}/\nu$ & 269/16 & 23/19 & 54/10 & 26/6\\
\hline
\multicolumn{5}{c}{Power-law} \\
\hline
$A_{V}$ & $7.1\pm 0.6$ & $11.6^{+0.8}_{-0.7}$ & $1.3^{+0.6}_{-0.5}$ & $<$0.96\\
$\alpha$\footnote{The power-law index, defined according to $F_{\nu}\propto \nu^{-\alpha}$.} & $-0.73\pm 0.13$ & $-2.29^{+0.24}_{-0.28}$ & $-0.36^{+0.17}_{-0.18}$ & $0.17^{+0.12}_{-0.47}$\\
$N_{\rm pl}$\footnote{The power-law normalization, corresponding to the flux in Jy at a reference frequency of $10^{14}$ Hz.} & $0.030\pm 0.003$ & $0.27\pm 0.03$ & $0.00191\pm 0.00015$ & $0.0039\pm 0.0006$\\
$\chi^{2}/\nu$ & 300/16 & 19/19 & 12/10 & 9/6\\
\hline
\multicolumn{5}{c}{2 Blackbodies} \\
\hline
$A_{V,1}$ & $17.5^{+0.8}_{-2.3}$ & -- & -- & --\\
$T_{1}$ (K) & $>$9400 & -- & -- & --\\
$R_{1}/d$ (\Rsun/kpc) & $<$4.36 & -- & -- & --\\
$A_{V,2}$ & $<$4.89 & -- & -- & --\\
$T_{2}$ (K) & $>$3800 & -- & -- & --\\
$R_{2}/d$ (\Rsun/kpc) & $<$1.13 & -- & -- & --\\
$\chi^{2}/\nu$ & 12/14 & 23/17 & 8/7 & 7/3\\
\hline
\multicolumn{5}{c}{Blackbody+Power-law} \\
\hline
$A_{V,bb}$ & $17.5^{+0.8}_{-2.1}$ & -- & -- & --\\
$T$ (K) & $>$9500 & -- & -- & --\\
$R/d$ (\Rsun/kpc) & $<$4.25 & -- & -- & --\\
$A_{V,pl}$ & $4.3^{+2.8}_{-1.1}$ & -- & -- & --\\
$\alpha^{a}$ & $-1.7^{+1.0}_{-2.9}$ & -- & -- & --\\
$N_{\rm pl}$$^{b}$ & $0.0015^{+0.0020}_{-0.0013}$ & -- & -- & --\\
$\chi^{2}/\nu$ & 12/14 & 18/17 & 8/7 & 5/3\\
\hline
\end{tabular}
\end{minipage}
\end{table}

\begin{table}
\caption{Summary of Results\label{tab:summary}}
\begin{minipage}{\linewidth}
\footnotesize
\begin{tabular}{cccccc} \hline \hline
IGR Name & $d$ (kpc) & $d_{\rm fiducial}$ (kpc) & $L$ (0.3--10 keV)\footnote{The X-ray luminosity in ergs~s$^{-1}$ measured by {\em Chandra} at the fiducial distance.} & $L$ (20--50 keV)\footnote{The X-ray luminosity in ergs~s$^{-1}$ measured by {\em INTEGRAL} at the fiducial distance.} & Spectral Type\\ \hline \hline
J16195--4945 & $\sim$3--15 & 5 (Norma-Cygnus?) & $1.4\times 10^{34}$ & $5.8\times 10^{34}$ & OBV or supergiant\\
J16207--5129 & $\sim$3--10 & 5 (Norma-Cygnus?) & $1.3\times 10^{35}$ & $8.3\times 10^{34}$ & OBV or supergiant\\
J16167--4957 & 1.8--4.2 & 3 (Scutum-Crux?) & $4.7\times 10^{34}$ & $1.7\times 10^{34}$ & later than A0V\\
J17195--4100 & $<$2.6 & $<$2.6 & $<$$2.0\times 10^{34}$ & $<$$1.6\times 10^{34}$ & later than A3V\\ \hline
\end{tabular}
\end{minipage}
\end{table}

\end{document}